\documentclass[epjCONF, twocolumn]{svjour}
\usepackage{graphicx}
\usepackage[varg]{txfonts} 
\usepackage[latin1]{inputenc}
\session-title{Assembling the puzzle of the Milky Way}
\begin{document}
\title{The dynamical fingerprint of gas-expulsion: Insights into the assembly of the Milky Ways' old GC system}
\author{Marks, Michael\inst{1,2}\fnmsep\thanks{Member of the International Max Planck Research School (IMPRS) for Astronomy and Astrophysics at the Universities of Bonn and Cologne, \email{mmarks@astro.uni-bonn.de}} \and Kroupa, Pavel\inst{1}}
\institute{Argelander-Institut f\"ur Astronomie, Universit\"at Bonn, Auf dem H\"ugel 71, D-53121 Bonn, Germany \and Max-Planck-Institut f\"ur Radioastronomie, Auf dem H\"ugel 69, D-53121 Bonn, Germany}
\abstract{
Since the oldest globular clusters (GCs) are early residuals from the formation of the Milky Way (MW), GCs were exposed to the likely evolving potential of our Galaxy. The expulsion of the residual-gas from the GC's embedded progenitors is sensitive to the conditions in the pre-MW gas cloud. By means of N-body computations it is shown that gas throw-out from initially mass-segregated GCs affect the shape of the low-mass stellar mass function (MF) and that its imprint might still be visible in the present-day MF (PDMF). The strength of the tidal-field at birth influences the degree of gas-expulsion driven low-mass-star depletion and therefore the PDMF probes the MW potential at the time of GC formation. It is argued that among the old GC population in the MW, younger GCs show stronger low-mass-star loss than older GCs. This is shown to be consistent with a contracting and self-gravitating cloud in which fluctuations in the pre-MW potential grow with time. An initially relatively smooth tidal field evolved into a grainy potential within a dynamical time-scale of the collapsing cloud (based on \cite{Marks2010}).
} 
\maketitle
\section{Introduction}
\label{sec:intro}
\begin{figure*}
\centering\resizebox{1.75\columnwidth}{!}{ \includegraphics{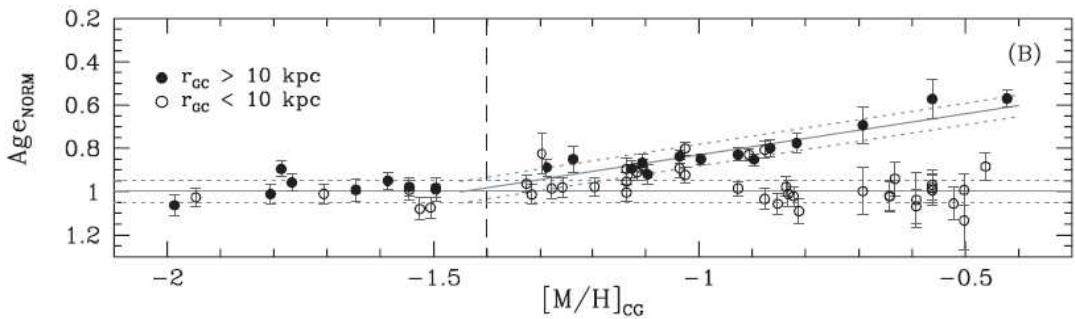} }
\caption{Distribution of GC relative ages as a function of metallicity (taken from \cite{MarinFranch2009}). Open and filled circles denote GCs with Galactocentric distance smaller and larger than $10$~kpc, respectively. Clusters group into coeval old GCs (relative age of $1$ corresponds to $\approx12.8$~Gyr) and the younger GCs following an age metallicity relation. The age dispersion in the old group is $\approx0.6$~Gyr, corresponding to a free-fall timescale (Sec.~\ref{sec:intro}). Old GCs witnessed the formation of the MW.}
\label{fig:ages}       
\end{figure*}
GCs of the MW may be categorized into old and young halo GCs (Fig.~\ref{fig:ages}). The latter group follows an age-metallicity relation and is thought to have an accretion origin \cite{SearleZinn1978}. The old group of GCs is roughly coeval, spanning a metallicity range from [M/H]$=-2$ to $-0.5$. The age dispersion of only $\approx600$~Myr, comparable to the free-fall time of a homogeneous sphere with the mass and size of the MWs dark matter halo ($1.9\times10^{12}M_{\odot},\;170$~kpc, \cite{WilkinsonEvans1999}), is ``not in contradiction with the formation from a single proto-system'' \cite{MarinFranch2009}, which formed the proto-MW probably along the lines originally suggested by Eggen et al. \cite{Eggen1962}. Therefore the MW GC halo consists of an in-situ (old GCs) and an accretion dominated, or more probably a tidally generated \cite{Pawlowski2011}, halo (younger GCs) with a radial transition between the two at about $\approx10$~kpc (see also \cite{Zolotov2009}). The old and coeval GCs thus witnessed the formation of the proto-MW.

In order to access the ambient conditions prevalent at GC birth we use the early violent residual-gas expulsion process which we show to leave an imprint in the PDMF. This \emph{dynamical fingerprint} depends on the strength of the tidal-field imposed on the GCs by the proto-MW which allows us, in combination with information about the metallicity (age) of the GCs, to trace back the very early events that contributed to the formation of the MW.   

\section{Gas expulsion and the PDMF}
\label{sec:pdmf}
\begin{figure*}
\centering\resizebox{1.75\columnwidth}{!}{ $\begin{array}{cc} \includegraphics{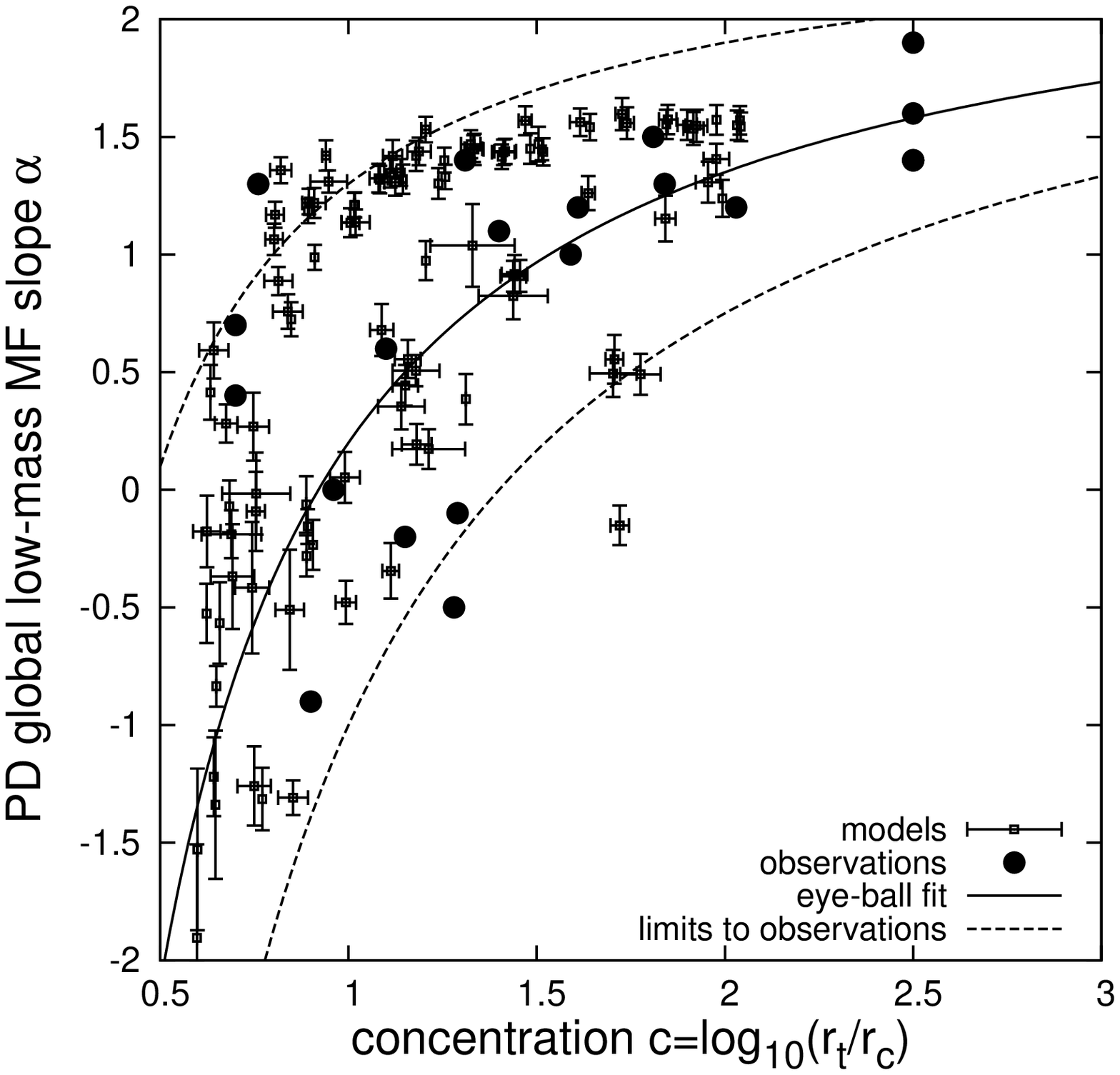} & \includegraphics{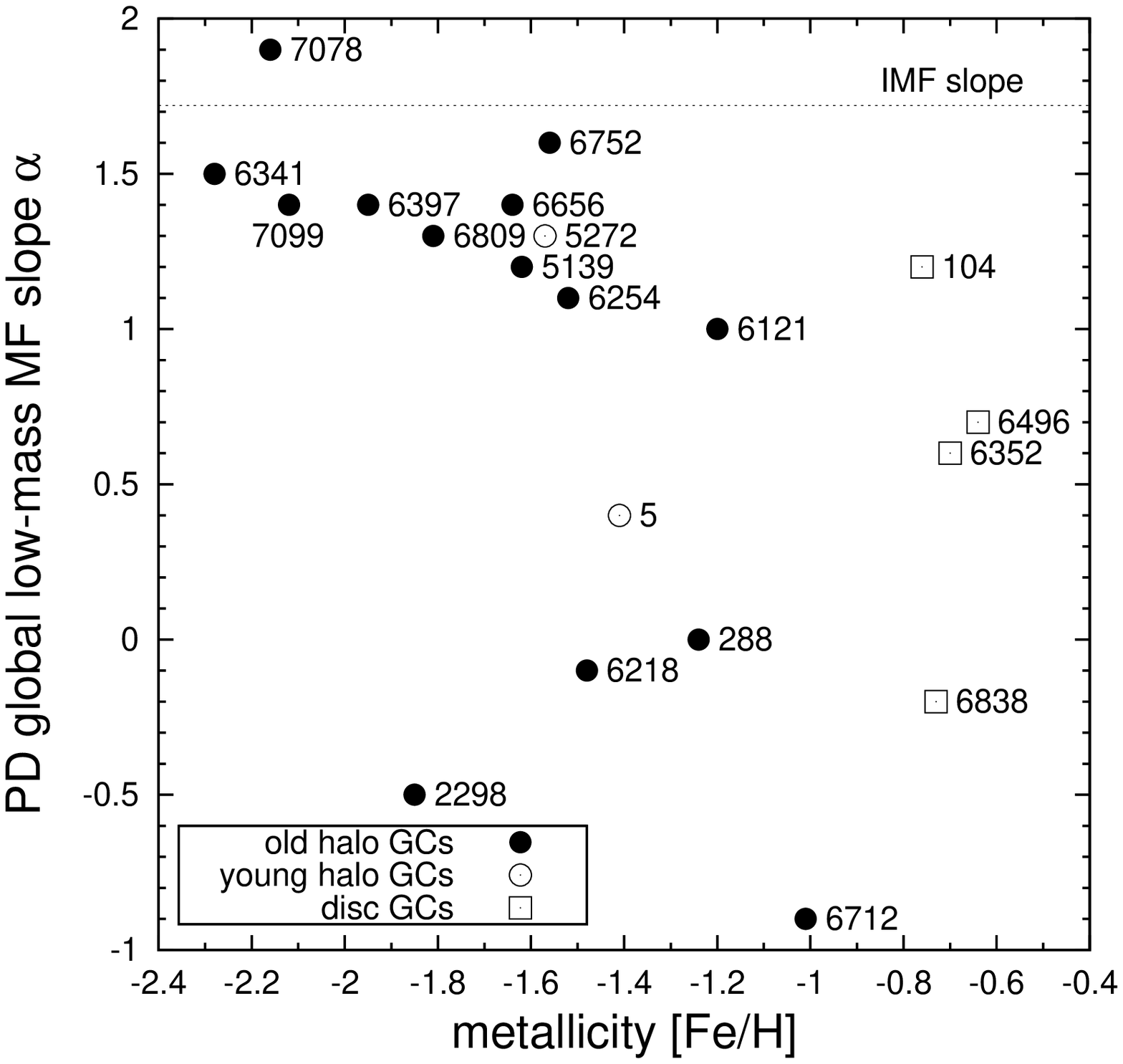} \end{array}$ }
\caption{\emph{Left:} Concentration parameter, $c=\log_{10}\left(r_t/r_c\right)$, vs. low-mass stellar PDMF slope, $\alpha$, ($dN/dm\propto m^{-\alpha}$) in the mass-range $0.3-0.8\;M_{\odot}$. Weakly concentrated clusters are strongly depleted in low-mass stars and no cluster with a high concentration and a depleted MF is found (filled dots). This trend (black solid line) can't be understood in terms of purely secular dynamical evolution. However, $N$-body integrations of mass-segregated clusters at the time of the emergence from their birth molecular cloud (squares with error bars) reasonably reproduce the observed trend within the observational limits (dashed lines). \emph{Right:} Global cluster metallicity, [Fe/H], vs. low-mass MF slope, $\alpha$. Data points are labelled with the respective NGC or Pal catalogue number of the GC. Clusters having a larger metallicity ($[{\rm Fe/H}]\gtrsim-1.5$) are depleted in low-mass stars ($\alpha\lesssim1$) giving support to the metallicity-dependent gas expulsion scenario (Sec.~\ref{sec:pdmf}).}
\label{fig:deMarchi}       
\end{figure*}
Classically, computations of star clusters start with an invariant IMF and the evolution of their stellar mass-function (MF) is determined by the evaporation of low-mass stars over the tidal-boundary through two-body relaxation \cite{Vesperini1997,BaumgardtMakino2003}. However, since clusters are observed to be segregated by stellar mass already at very young ages, frequently interpreted as evidence for primordial mass-segregation, it may be questioned whether this is the whole story. Upon gas-expulsion and the subsequent expansion of the cluster, stars in the cluster outskirts are stripped off the cluster through the external tidal-field. If low-mass stars form at the cluster periphery, the low-mass star population will be preferentially depleted, leading to a flattening of the mass-function before classical evolution starts.

Marks et al. \cite{Marks2008} showed this to be the case by making use of the grid of residual-gas expulsion models of Baumgardt \& Kroupa \cite{Baumgardt2007}. If clusters expell their gas on a short time-scale, the stars in the cluster, assumed to be close to virial equilibrium before gas throw-out starts, cannot adjust quickly enough to the changing potential, hence leading to strong cluster expansion and effective loss of stars from the cluster outskirts. The effect of mass-loss over the tidal-boundary is enhanced if the star formation efficiency is low, i.e. a significant fraction of the total mass remains in the form of gas which is to be expelled, and if the tidal-field is strong.

Marks et al. \cite{Marks2008} show that if the cluster is mass-segregated, low-mass stars will be removed at first, hence flattening the low-mass stellar MF. If the expansion and tidal-field (and, thus, mass-loss) are strong, the concentration of the cluster decreases in parallel with the depletion of the low-mass star population in a mass-segregated cluster. But if expansion and the tidal-field (and mass-loss) are weak, the concentration remains close to the initial value and the MF is only weakly depleted of low-mass stars.

This has indeed been observed by de Marchi et al. (\cite{deMarchi2007}, Fig.~\ref{fig:deMarchi}). That gas-expulsion is indeed at work in initiating this trend is strengthened by Marks \& Kroupa \cite{Marks2010}, who show higher-metallicity GCs to have flatter present-day mass-functions (PDMFs). Although the observed trend had been recognized earlier \cite{McClure1986,Djorgovski1993} it lacked an explanation until recently. In particular, this trend is difficult to reconcile with standard dynamical evolution scenarios with a universal IMF as it is unclear how dynamics could possibly know about the metal content of a cluster. Marks \& Kroupa \cite{Marks2010} interpret the PDMF trend with metallicity as evidence for metal-dependent cluster winds in initially mass-segregated clusters: As radiation couples better to metal-rich gas, the expulsion of the left-over gas from star formation is quicker and perhaps more efficient, leading to strong expansion of the cluster as a whole and thereby loosing preferentially low-mass stars over the tidal boundary.

The de Marchi trend \cite{deMarchi2007} was also unexpected from a classical point of view but is a natural outcome of gas expulsion. Although it has been shown that the depletion of low-mass stars occurs quicker if clusters start mass-segregated initially \cite{Baumgardt2008}, such classical scenarios without gas lack an understanding of the metallicity trend.

\subsection{Assembly of the old halo of GCs}
\label{sec:assembly}
\begin{figure}
\centering\resizebox{1.00\columnwidth}{!}{ \includegraphics{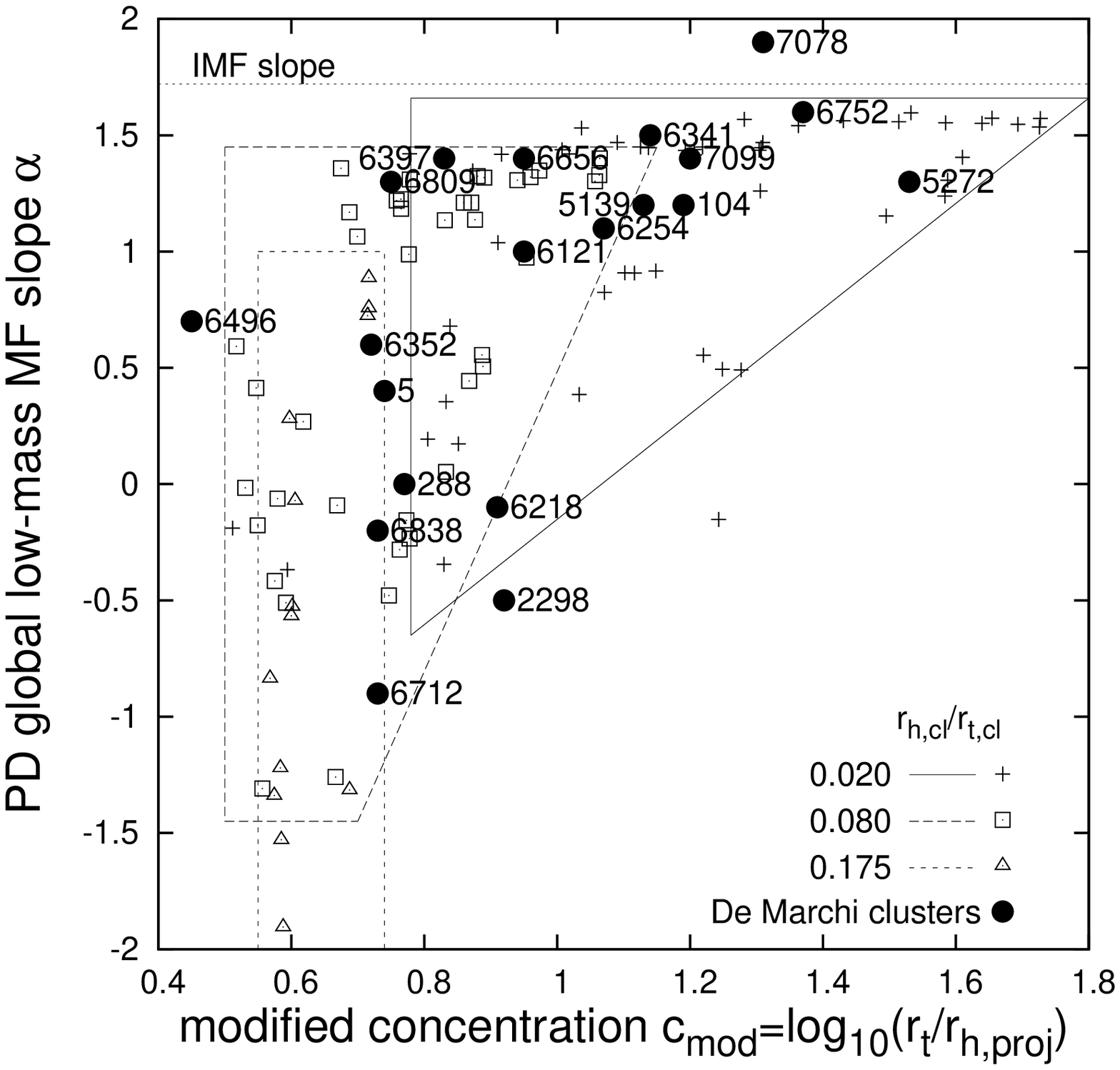} }
\caption{As in Fig.~\ref{fig:deMarchi} (left) but with a modified concentration parameter (see \cite{Marks2010}) on the abscissa. The theoretical data are coded for the initial tidal-field strength, in the simulations of \cite{Baumgardt2007} with different symbols. Open symbols and crosses depict the $N$-body results, the filled circles labelled with their respective catalogue number are the data from the de Marchi GC sample. Quadrangles and triangles show areas with similar tidal-field strengths. All initial models started from the dashed line, which is the slope of the canonical IMF \cite{Kroupa2001} in the mass-range $0.3-0.8\;M_{\odot}$ ($\alpha\approx1.72$), in which the MF slope of the observed GCs was measured. The tidal-field strength increases with decreasing concentration and increasing low-mass star depletion. Thus, clusters which experienced stronger tidal-fields are more metal-rich (i.e. younger, Fig.~\ref{fig:deMarchi}, right panel).}
\label{fig:tidalfields}       
\end{figure}
\begin{figure*}
\centering\resizebox{1.75\columnwidth}{!}{ \includegraphics{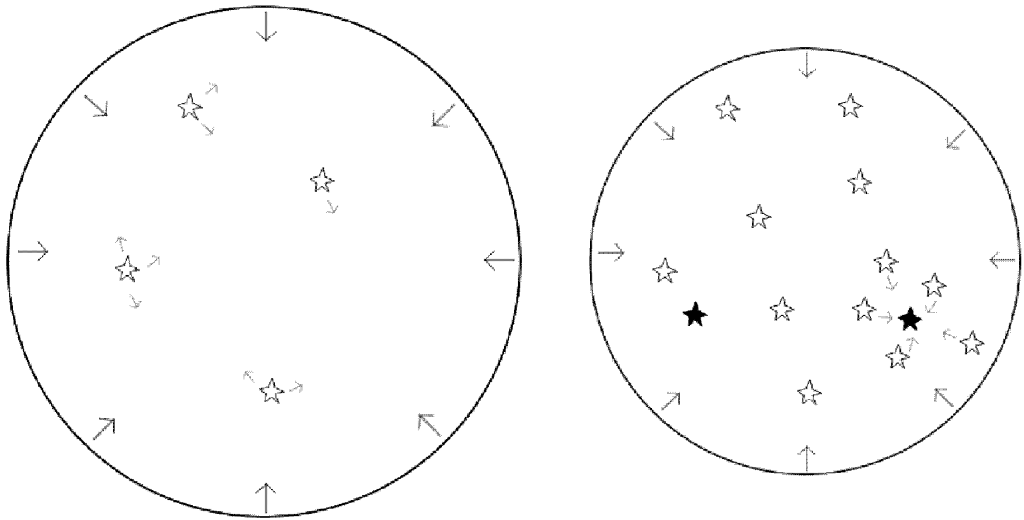} }
\caption{Contraction scenario as described in Sec. \ref{sec:assembly}. \textit{Left-hand picture:} In a collapsing cloud the first clusters (the stars in the picture) form all over the cloud in a smooth potential. The conditions are similar for all of them leading to comparable MF slopes and concentrations after gas expulsion. These clusters enrich their immediate environment with metals. \textit{Right-hand picture:} As the collapse proceeds more clusters form and the potential becomes grainy enhancing the mean tidal-field strength. Star clusters, e.g. the right filled cluster, forming near other clusters or dense clouds experience extreme conditions. They are more strongly enriched in metals from the surrounding objects than other clusters, e.g. the left filled cluster which is located in a more isolated place of the grainy potential, and they experience stronger tidal-fields than the first-forming clusters. This leads to strong differences between the formation sites and, in turn, to variations in the PDMF slope.}
\label{fig:assembly}       
\end{figure*}
From the models it is found that as the concentration decreases, the tidal-field strength increases, i.e. present-day low-concentration clusters have formed in stronger tidal-fields (Fig.~\ref{fig:tidalfields}). Furthermore the GCs of the old halo appear to be co-eval to a good approximation (Sec.~\ref{sec:intro}). Despite their formation at the same time the clusters show a large spread in metallicity ($-2.4 < [Fe/H] < -1.0$ for the old halo GCs in \cite{deMarchi2007}). If the metal content of the more metal-rich old population GCs comes from the products of stellar evolution which have been produced by earlier forming massive stars in other clusters, the inter-cluster medium (ICM) will be successively enriched with time. This material will then be recycled into later forming clusters. In this case a higher metallicity as observed for the low-$\alpha$ GCs (Fig.~\ref{fig:deMarchi}) suggests that these clusters formed from an ICM enriched in metals and they should thus be somewhat younger than the relatively more metal-poor clusters. The age difference can then be at most a few hundred Myr only, corresponding to the free-fall time-scale (Sec.~\ref{sec:intro}).

These ideas lead us to a picture of the formation of the old inner and co-eval Galactic GC system, which might have formed during the contraction of the pre-MW gas cloud out of which finally, after an additional long period of ongoing accretion and merging of formerly extragalactic systems \cite{SearleZinn1978}, the MW has emerged (possibly along the lines originally proposed by Eggen et al. \cite{Eggen1962}). These ideas are summarised in Fig.~\ref{fig:assembly}.

In this frame, clusters which are depleted of low-mass stars, i.e. those that reside in stronger tidal-fields were born at a later stage of galaxy formation. The environment in which the clusters formed must then have changed drastically within a short time. The initial conditions can be understood if the overall potential was rather smooth in the beginning. The first GCs formed all over the cloud experiencing similar smooth tidal-fields more or less independent of Galactocentric distance and explaining the comparable strengths of the tidal-fields. These first clusters enriched the ICM of their local environment with metals, from which the somewhat younger clusters of the old population were born.

The pre-galaxy gas cloud contracted due to self-gravitation during this process. The cloud may have become clumpy and substructures emerged. Fragmentation of the cloud into massive star cluster forming regions made the potential grainy, thereby explaining the different and on average stronger tidal-field strengths among the younger clusters. Clusters forming next to other massive, dense objects are forced to expel their gas under more extreme conditions (a stronger tidal-field) than expected from their present-day Galactocentric distance.

Gas expulsion from early forming clusters may trigger the formation of new clusters (via gas compression) in their immediate surrounding perhaps leading to cluster complexes, i.e. clusters of star clusters. It is imaginable that similar processes were at work in the GC forming gas cloud, increasing the average local tidal-field strengths.

These effects lead to depleted MFs and lower concentrations after gas expulsion in the younger of the old halo GCs. In this sense the history of events that lead to the inner GC system may involve local rapid (on time-scales of hundreds of Myr) re-arrangements of the interstellar matter superposed on the overall contraction or collapse to the final population II spheroid.

\section*{Acknowledgements}
MM was supported for this research through a stipend from the International Max Planck Research School (IMPRS) for Astronomy and Astrophysics at the Universities of Bonn and Cologne.

\end{document}